\documentclass[sigconf]{acmart}
\AtBeginDocument{%
  }

\setcopyright{acmlicensed}
 \copyrightyear{2025} 
\acmYear{2025} 
\setcopyright{cc}
 \setcctype{by-nc-nd}
 \acmConference[GLSVLSI '25]{Great Lakes Symposium on VLSI 2025}{June
 30-July 2, 2025}{New Orleans, LA, USA}
 \acmBooktitle{Great Lakes Symposium on VLSI 2025 (GLSVLSI '25), June 30-July 2, 2025, New Orleans, LA, USA}
 \acmDOI{10.1145/3716368.3735166}
 \acmISBN{979-8-4007-1496-2/2025/06}

\usepackage{multirow}

\usepackage{subcaption} 

\begin{document}

\title{Learning Cache Coherence Traffic for NoC Routing Design}


\author{Guochu Xiong}

\affiliation{%
  \institution{College of Computing and Data Science, Nanyang Technological University}
  \country{Singapore}
}
\email{guochu.xiong@ntu.edu.sg}

\author{Xiangzhong Luo}
\affiliation{%
  \institution{College of Computing and Data Science, Nanyang Technological University}
  \country{Singapore}
}
\email{xiangzhong.luo@ntu.edu.sg}

\author{Weichen Liu}
\authornote{Corresponding author}
\affiliation{%
  \institution{College of Computing and Data Science, Nanyang Technological University}
  \country{Singapore}
}
\email{liu@ntu.edu.sg}
 



\begin{abstract}
  The rapid growth of multi-core systems highlights the need for efficient Network-on-Chip (NoC) design to ensure seamless communication. Cache coherence, essential for data consistency, substantially reduces task computation time by enabling data sharing among caches. As a result, routing serves two roles: facilitating data sharing (influenced by topology) and managing NoC-level communication. However, cache coherence is often overlooked in routing, causing mismatches between design expectations and evaluation outcomes. Two main challenges are the lack of specialized tools to assess cache coherence’s impact and the neglect of topology selection in routing. In this work, we propose a cache coherence–aware routing approach with integrated topology selection, guided by our Cache Coherence Traffic Analyzer (CCTA). Our method achieves up to 10.52\% lower packet latency, 55.51\% faster execution time, and 49.02\% total energy savings, underscoring the critical role of cache coherence in NoC design and enabling effective co-design.
\end{abstract}





\maketitle

\section{Introduction}
Advancements in computational power increasingly rely on multi-core system design, where efficient communication is essential. Network-on-Chip (NoC) serves as the backbone of data transfer and coordination, ensuring optimal performance and scalability. A well-designed NoC enhances throughput and minimizes latency, directly boosting computational efficiency. Over time, routing has emerged as a critical research area, bridging NoC architecture and communication efficiency while driving major breakthroughs \cite{b8,b24,b25,b30,b35}. With AI-driven advancements, routing strategies have significantly improved dynamic NoC management and performance optimization, reinforcing their role in innovation.

In parallel, cache coherence is essential for multi-core communication, especially in NoC systems, where it ensures a consistent memory view across processors. Core requests generate coherence messages to update states or fetch data, enabling efficient parallel access via protocols like MSI, MESI, and MOESI \cite{b12}. Research has optimized coherence protocols to reduce traffic \cite{b13, b14}, improve energy efficiency \cite{b15}, and minimize latency \cite{b16}. However, existing tools overlook cache coherence performance, focusing only on NoC or system-level metrics, making it difficult to evaluate coherence optimizations and their impact on NoC design. This gap hinders a comprehensive assessment of their true effectiveness.

\begin{figure}[htbp]
    \centering
    
    \begin{subfigure}[b]{0.45\textwidth}
        \centering
        \includegraphics[width=\linewidth]{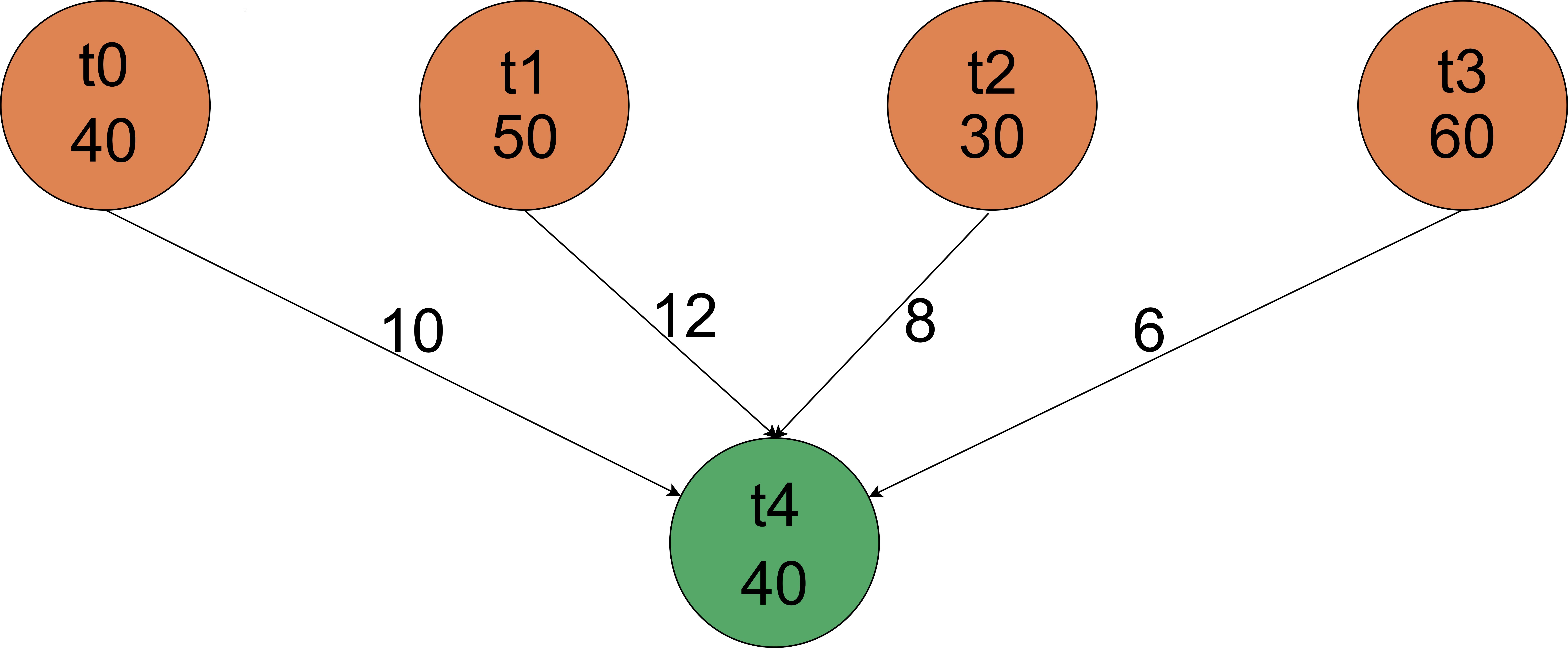}
        \caption{An example of task graph}
        \label{fig:task graph}
    \end{subfigure}

    \begin{subfigure}[b]{0.45\textwidth}
        \centering
        \includegraphics[width=\linewidth]{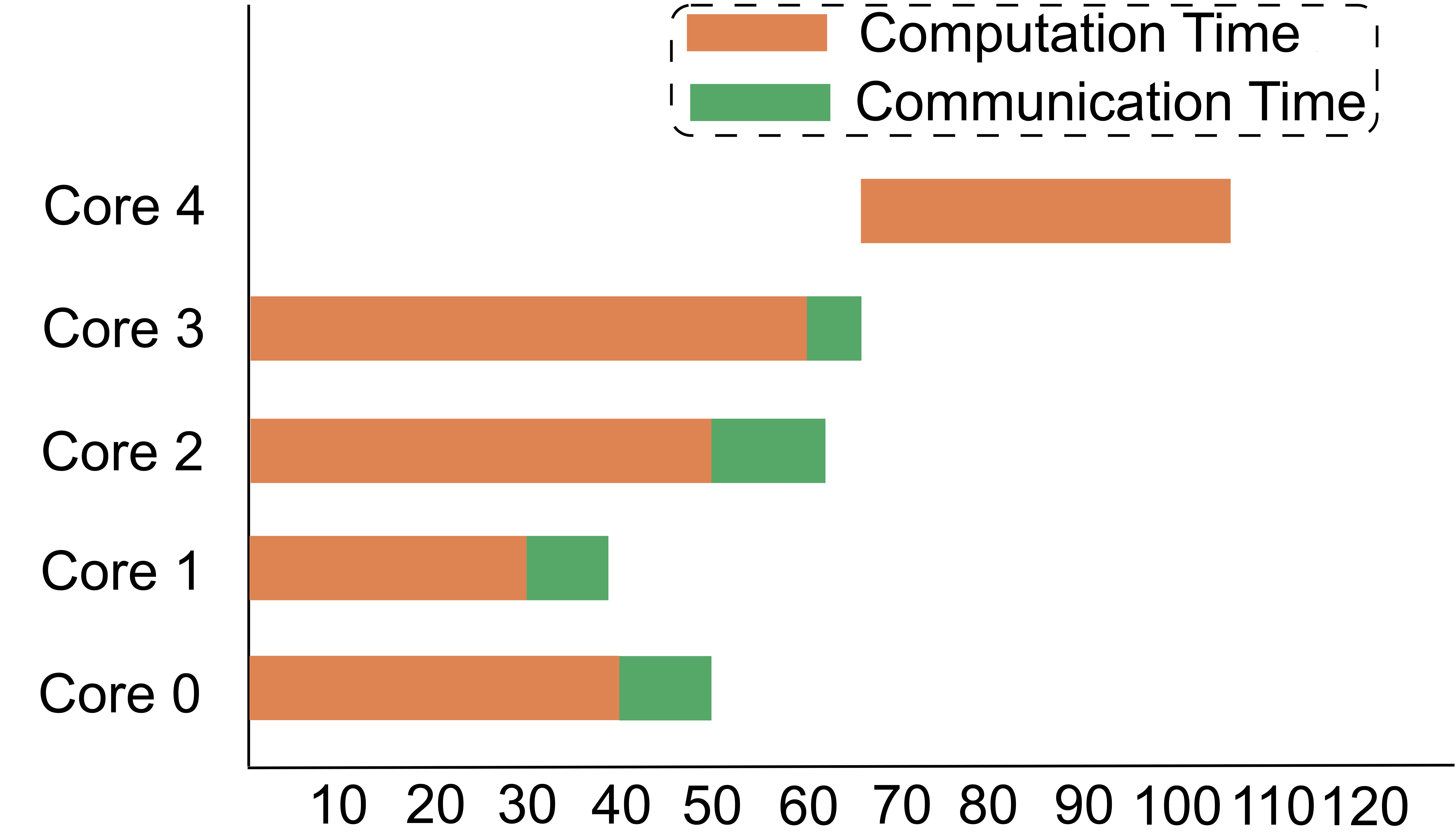}
        \caption{Synthetic traffic with optimal mapping and routing}
        \label{fig:NoC traffic}
    \end{subfigure}  
    
    \begin{subfigure}[b]{0.47\textwidth}
        \centering
        \includegraphics[width=\linewidth]{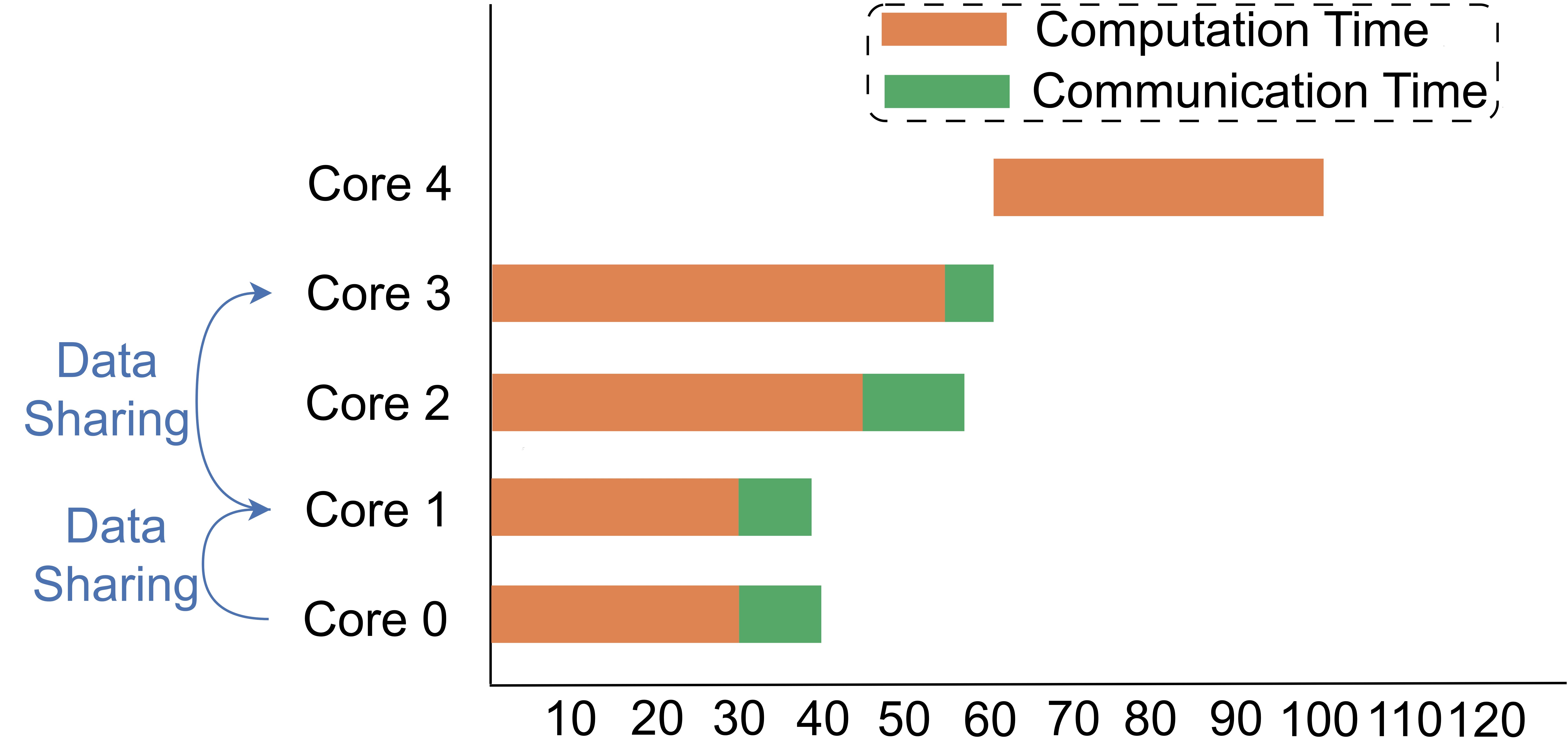}
        \caption{Cache coherence traffic with optimal mapping and routing}
        \label{fig:cache coherence traffic}
    \end{subfigure}  
      
    \caption{Motivation example: (a) DAG Task Graph Representation; (b) Synthetic traffic without cache coherence protocol; (c) Cache coherence traffic with data sharing between cores.}
    \label{fig:motivation_1}
\end{figure}

Despite its crucial role in NoC communication, cache coherence is often overlooked in NoC design. Consider the task graph in Figure \ref{fig:task graph}, where the number of cores matches the number of tasks. Assume Task $t_0$ and $t_1$ share some data initially stored in core 0, while $t_1$ and $t_3$ share other data stored in core 3. Most NoC designs \cite{b19, b37, b38}rely on synthetic traffic (Figure \ref{fig:NoC traffic}), which ignores cache coherence. However, cache coherence traffic (Figure \ref{fig:cache coherence traffic}) better reflects real-world applications, reducing computation time even with optimal mapping and routing. This improvement comes from direct cache-to-cache data sharing, avoiding the costly process of fetching data from main memory. Consequently, routing plays a dual role: (1) facilitating efficient data sharing during task computation through cache-to-cache communication (this portion of communication time is included in the computation time), and (2) managing purely NoC-level communication, counted as communication time. While topology influences mapping and routing effectiveness, optimizing communication time also requires accounting for congestion and network dynamics.

One persistent challenge is integrating topology selection as a core component of routing decisions. Most approaches either fix topologies\cite{b30,b35} or treat them as part of the environment \cite{b8,b23,b24,b25}, often leading to suboptimal outcomes, as routing performance depends heavily on network structure, ignoring topology impact can significantly undermine a routing strategy’s effectiveness.

\begin{figure}
    \centering
    \includegraphics[width=\linewidth]{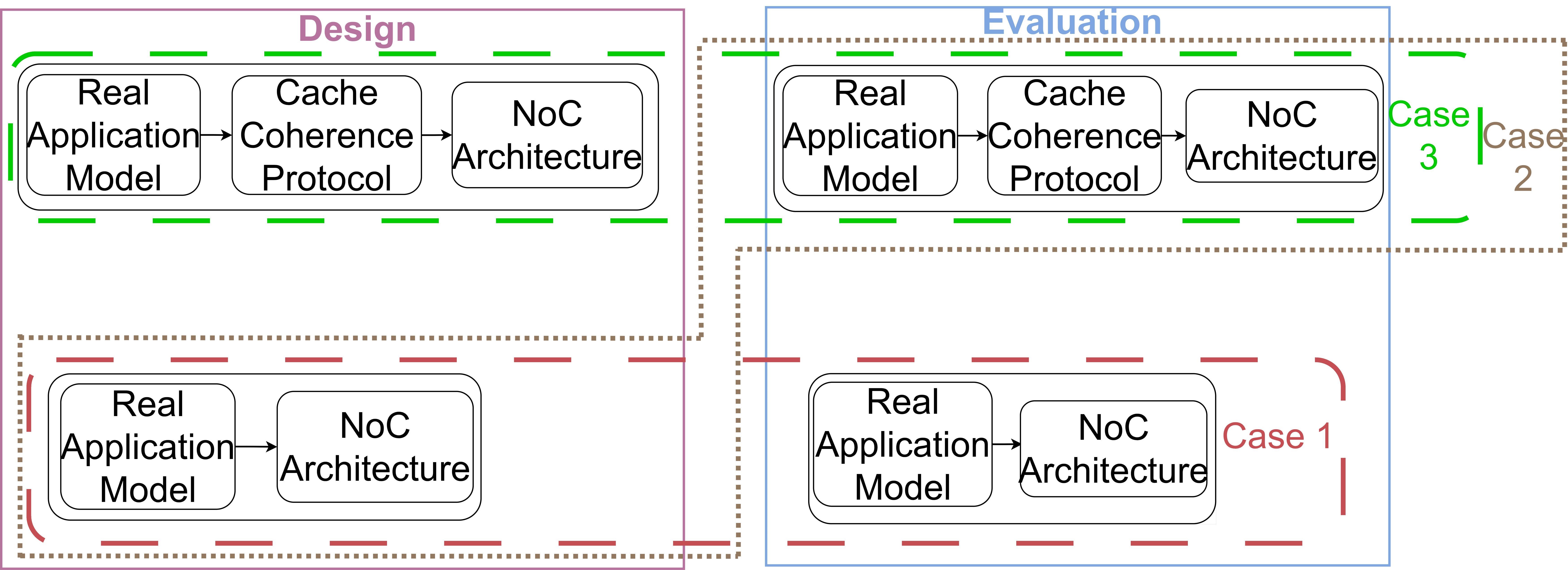}
    \caption{Difference of NoC design workflow. }
    \label{fig:environment}
\end{figure}

Additionally, current routing designs overlook cache-level data sharing from cache coherence, focusing solely on the communication within NoC. As shown in Figure \ref{fig:environment}, existing routing strategies generally fall into two categories \cite{b2, b8, b24, b25, b26, b29, b30}: The first focuses solely on design and evaluation within synthetic traffic (Case 1), while the second designs using synthetic traffic but evaluates the system with cache coherence present (Case 2). In Case 1, excluding cache coherence fails to capture real-world traffic overhead, while in Case 2, evaluating performance with cache coherence after the design phase misses opportunities to optimize routing from the outset. Cache coherence adds significant traffic, increasing data transfer demands and straining the network. As core counts scale, challenges like energy consumption and congestion intensify, further impacting communication efficiency between caches. Ignoring cache coherence during design creates a mismatch between expected and actual performance. To address these shortcomings, Case 3 integrates cache coherence into both design and evaluation. Coherence traffic alters communication patterns by increasing cache-to-cache exchanges, adding overhead and potentially congesting conventional routing paths. By incorporating cache coherence from the outset, routing strategies can manage traffic more effectively, mitigating congestion and improving NoC performance.

In this paper, we highlight three key observations: (i) NoC design overlooks cache coherence-induced data sharing in task computation, (ii) design objectives often misalign with actual performance when cache coherence is considered, and (iii) a lack of tools prevents a deeper understanding of cache coherence's impact on NoC design. To address these gaps, we introduce a novel analysis tool fully integrated with Gem5 \cite{b17} to evaluate cache coherence performance in realistic scenarios. Additionally, we propose a DRL-based approach that optimally selects topology and routing paths, dynamically adapting to network structure while accounting for cache coherence traffic, resulting in improved real-world performance. Our contributions include: (i) developing the Cache Coherence Traffic Analyzer (CCTA) to evaluate key metrics and provide insights into cache coherence effects, (ii) integrating cache coherence features into our framework, highlighting its critical role in NoC design and enabling new co-design opportunities, (iii) introducing a DRL-based framework that optimizes topology and routing by leveraging the interplay between NoC traffic and cache coherence. During design, the DRL agent learns the environment, enabling selection of optimal topologies and routing paths. For users with fixed topologies, we provide pretrained models as baselines for further optimization, benchmarking, and validation. Both the framework and models are fully integrated into Gem5, offering a comprehensive solution for NoC design and evaluation, and (iv) demonstrating the superiority of our approach by consistently outperforming existing algorithms, particularly in handling cache coherence complexities, leading to enhanced performance and efficiency.

\section{Related Work}
To the best of our knowledge, this is the first work to propose cache-coherence-aware DRL-based approach that jointly optimizes topology selection and routing, bridging NoC design and cache coherence into a comprehensive solution.

NoC design has advanced significantly. For instance, Garnet\cite{b1} provides a cycle-accurate CMP evaluation model, while ArSMART NoC\cite{b19} enhances routing through clustering and dynamic transmission, and neural networks offer faster, more accurate latency estimation\cite{b20}. However, existing routing algorithms struggle to integrate topology selection with routing decisions. While RL-based routing is used to reduce congestion and improve performance\cite{b31,b32,b33} and system feedback aids low-latency designs\cite{b35}, multi-objective optimization remains confined to fixed topologies\cite{b26}. Some works treat topology as static\cite{b8,b30} or separate it from routing\cite{b23,b24,b25}, overlooking that topology selection should be guided by routing since the latter inherently depends on network structure.

Recent cache coherence advancements target energy reduction and traffic optimization. For example, \cite{b13} improves directory-based coherence by distinguishing private from shared data to lower traffic and DRAM accesses, while \cite{b14} introduces DiCo-CMP, which outperforms Token-CMP in traffic and area efficiency. However, most NoC designs treat cache coherence as an external factor, missing opportunities for integrated performance gains. Similarly, cache coherence analysis tools primarily focus on time measurement. For instance, \cite{b38} estimates worst-case execution time by integrating isolated cache analysis, timer-based retention, and time-coherence—albeit limited to snooping-based protocols and requiring modifications—while \cite{b39} gathers time statistics by altering classical protocols.

Therefore, there remains a pressing need to integrate cache coherence into NoC design and analysis while adapting topology to routing decisions. Our approach fills this gap by developing a cache coherence analysis tool, combining it with NoC metrics during routing design, and optimizing topology and routing holistically, ultimately leading to more efficient traffic distribution, reduced energy consumption, and enhanced system performance.

\section{Methodology}

\label{sec:method}
\begin{figure}
    \centering
    \includegraphics[width=\linewidth]{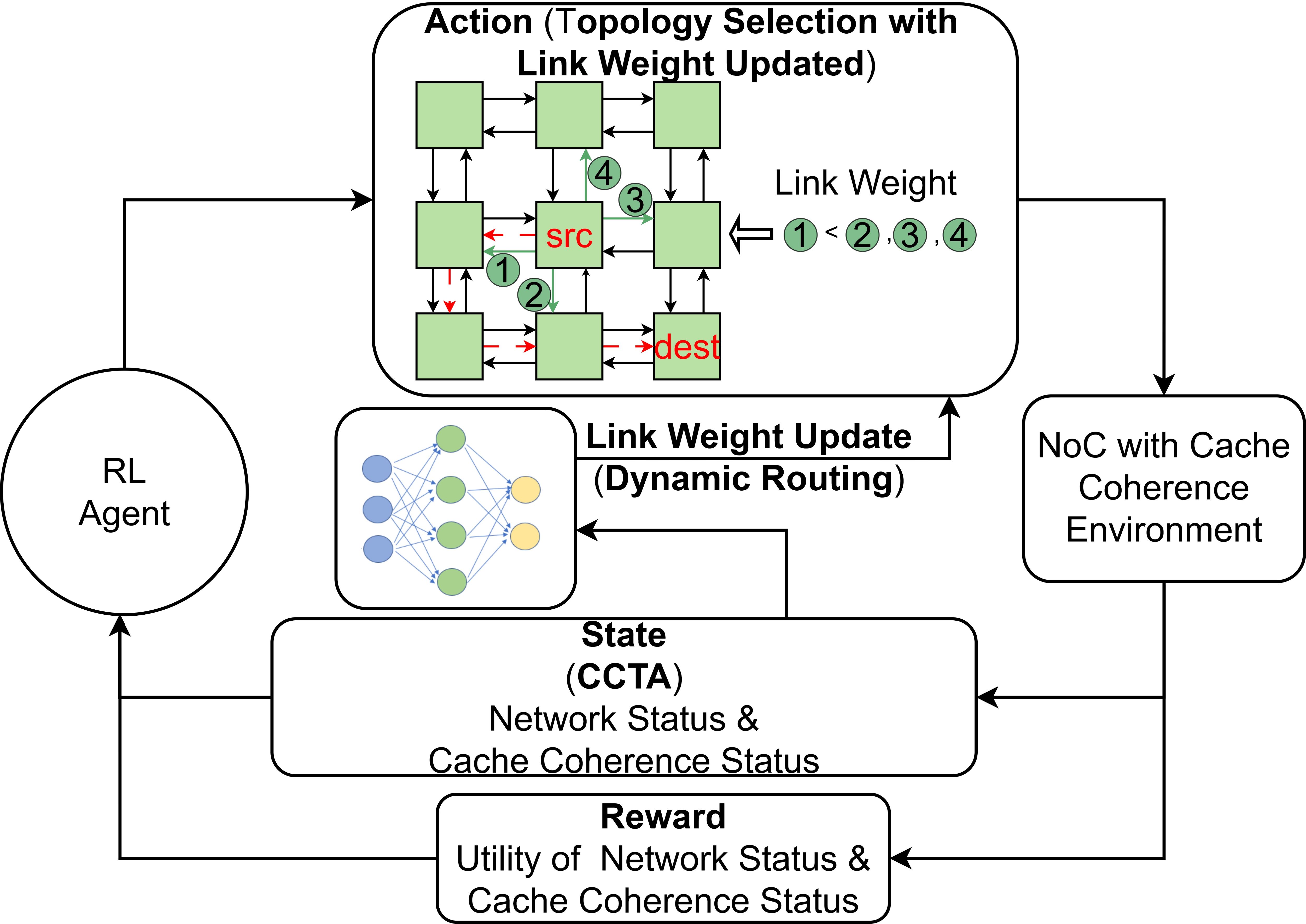}
    \caption{Proposed DRL-based topology selection with neural network learning for dynamic routing.}
    \label{fig:RL_routing}
\end{figure}

This section outlines our framework that integrates deep reinforcement learning (DRL) with cache coherence-aware decision-making for dynamic topology selection and routing in NoC systems. We focus on two main components: the Cache Coherence Traffic Analyzer (CCTA) for measuring and analyzing cache coherence metrics (Section \ref{cache coherence measurenment}), and the DRL-based topology selection with routing optimization model (Section \ref{Co-optimization}), as shown in Figure \ref{fig:RL_routing}.

\subsection{Cache Coherence Traffic Analysis}
\label{cache coherence measurenment}

Cache coherence maintains data consistency and shareability in multi-core systems. Before memory operations propagate, the protocol ensures data copies in other caches are updated or invalidated, making changes visible to all cores. Coherence messages, like invalidation or update requests, are sent across the NoC, contributing to traffic and affecting NoC performance and system efficiency.

\begin{figure}[h!]
    \centering
    \includegraphics[width=\linewidth]{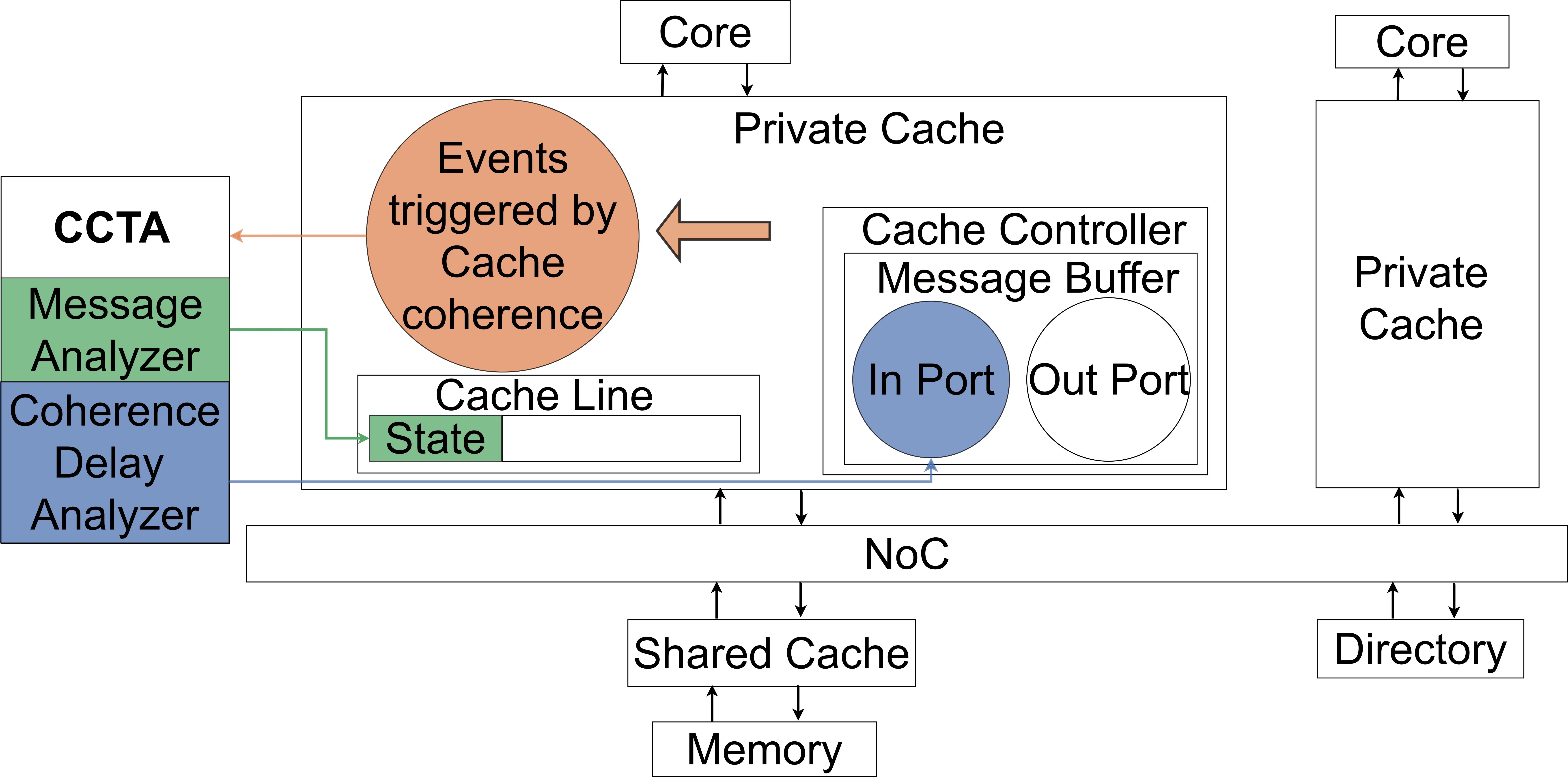}
    \caption{The framework of CCTA.}
    \label{fig:cache_coherence_measurement_tool} 
\end{figure}

Despite its crucial role in NoC and system performance, cache coherence introduces communication overhead that impacts packet transmission, making accurate measurement essential. However, existing methods lack dedicated evaluation of coherence traffic metrics in NoC design, hindering progress assessment and limiting verification of its impact. To address this gap, we developed the Cache Coherence Traffic Analyzer (CCTA) (Figure \ref{fig:cache_coherence_measurement_tool}).

\begin{figure}[!htbp]
    \centering
    \includegraphics[width=\linewidth]{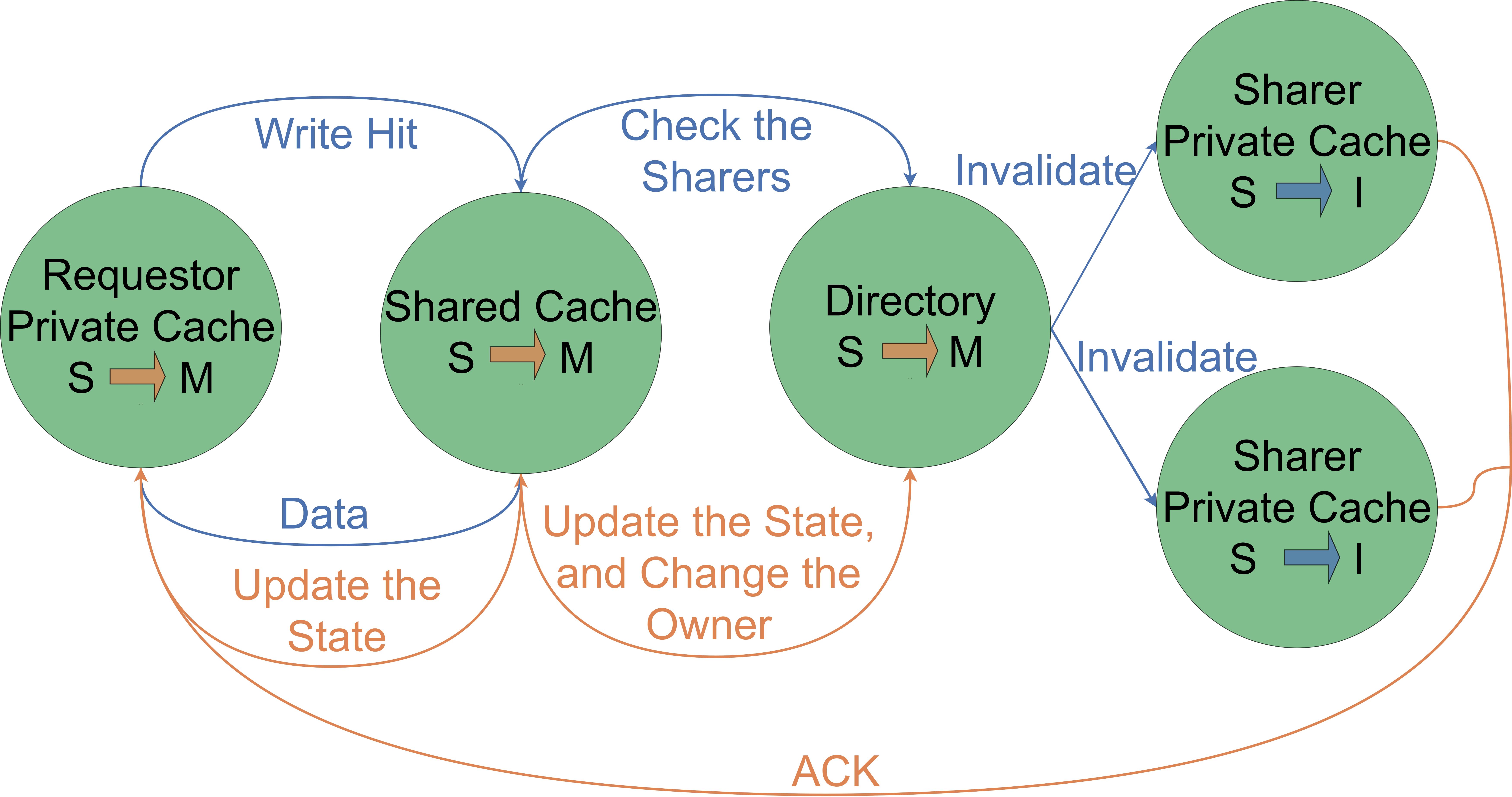}
    \caption{The process of cache coherence as private cache in S state in write hit case.}
    \label{fig:writehitS}
\end{figure}

Cache coherence events fall into three categories: write hits in private caches with an S (Shared) state, read misses, and write misses in private or shared caches. For example, in the MESI directory-based protocol \cite{b12}, the most widely used in current literature, when a private cache receives a write request in the S state, the cache coherence protocol initiates message transmissions between system components until the private cache controller receives acknowledgment (ACK) messages (shown in Figure \ref{fig:writehitS}). Without cache coherence (e.g., in synthetic traffic), these messages stay within the requestor, avoiding extra NoC traffic. Thus, cache coherence increases network traffic and CPU delay, which is the time from a write request in the S state to execution, entirely attributed to coherence overhead. Similarly, during read and write misses, the protocol manages message transmissions to maintain data consistency across caches.

The dynamic routing of diverse message types across multiple ports during NoC and cache coherence interactions makes metric collection and analysis challenging. Specifically, it is difficult to (i) identify cache coherence messages, (ii) determine their starting points, and (iii) manage communication across multiple caches and levels. Additionally, the complexity of cache coherence protocols, where a single state can generate multiple message types and the same message can traverse different states, further complicates classification and collection. CCTA addresses these challenges by using intermediate states to capture transient events alongside primary theoretical states in each cache.

To evaluate timing metrics (CPU delay, average write miss time (including CPU delay), and average memory fetch time), CCTA marks the start and end points within the L1 private cache where the initial request and final response occur, using primary and intermediate states. Timestamps (orange markers in Figure \ref{fig:cache_coherence_measurement_tool}) log each request’s duration, while monitors at private cache controller in-ports record the first and last message timestamps (blue markers in Figure \ref{fig:cache_coherence_measurement_tool}).  These intervals (orange and blue lines in Figure\ref{fig:writehitS}) are accumulated to capture the complete timeline of packet traversal through the NoC, from initiation to final state transitions.

CCTA analyzes cache coherence communication by tracking message transmissions across NoC. Each time a cache line’s state changes—from request initiation to completion—message collection is triggered and continues until the final transmission (green markers in Figure \ref{fig:cache_coherence_measurement_tool}). As shown in Figure \ref{fig:writehitS}, every state change in the requestor is recorded as a message transmission event, capturing all protocol-generated traffic. This process starts when the first message signals a state change (e.g., a write request for a Shared-state cache line) and ends with the final message (e.g., upon receiving an ACK). By integrating CCTA, we achieve precise measurement and in-depth analysis of cache coherence communication.

\subsection{Topology Selection and Routing Design}
\label{Co-optimization}
Our methodology leverages DRL to jointly optimize topology selection and routing in a dynamic NoC and cache coherence environment. We choose DRL over traditional RL, adaptive routing\cite{b4}, and genetic algorithms\cite{b3} because it learns multi-objective rewards from both NoC and cache coherence metrics while adapting in real time. Unlike RL (single-objective reward), genetic algorithms (slow convergence), or adaptive routing (complex congestion management), our dual-network architecture—featuring a Q-network combined with an$\epsilon$-greedy strategy for decision making and a WeightPredictor for fine-tuning simulation parameters, dynamically adjusts to network conditions, significantly enhancing overall adaptability and performance.

\textbf{States.} A unique aspect of our approach is merging cache coherence states (e.g., number of coherence messages, CPU delay, average write miss time) with NoC states (e.g., average packet latency, average packet delay, average link utilization) into a single model state space. This integrated representation enhances context-awareness, enabling more informed decisions that consider both network and cache coherence metrics. The DRL agent and link-weight learning process both use this comprehensive state representation as input.

\textbf{Action.} Our DRL framework jointly optimizes topology selection and link-weight learning. For topology selection, the action space consists of candidate topologies. A Q-network computes Q-values based on NoC and cache coherence metrics, and an $\epsilon$-greedy policy—where $\epsilon$ decays as 1⁄(episode+1)—chooses the highest-valued topology while still exploring alternatives. For example, if congestion is high, the agent may select a topology with more routing paths even if it’s not typically optimal. For routing, the Link Weight Update module employs a three-layer MLP (256 neurons per hidden layer, ReLU activations, and 50\% dropout) whose output layer dynamically adjusts to the chosen topology and core count. Despite its simplicity, this network continually trains in real time to select the lowest-cost paths based on current conditions. Meanwhile, Garnet supports bidirectional links to further enhance routing flexibility.

\textbf{Reward.} Unlike traditional methods that optimize latency or energy in isolation, our approach balances NoC and cache coherence trade-offs, minimizing average packet latency ($L_t$), CPU delay ($H_t$), average packet delay ($D_t$), and cache coherence messages ($C_t$). Normalization values are empirically chosen based on simulated metric ranges to ensure proportional contributions and prevent dominance by any single factor:
\begin{equation}
        R_t = -(\alpha_1 \times L_t + \alpha_2 \times H_t + \alpha_3 \times D_t +\alpha_4 \times C_t)
\end{equation}

\section{Experiment}
\label{sec: experiment}

The experiment, using the PARSEC 2.1 benchmark suite \cite{b27} in the Gem5 simulator, evaluates performance in real-world scenarios. Energy consumption is measured with McPAT \cite{b28}. Simulations are conducted on NoC configurations with 16 and 64 cores, as shown in Table~\ref{tab:setup}. The system employs a directory-based MESI cache coherence protocol, which is both well-studied in academia and aligned with industry standards such as the Coherent Hub Interface (CHI), ensuring realistic modeling of coherence interactions.


\begin{table}[h!]
\centering
\caption{Platform Parameters}
\renewcommand{\arraystretch}{1.2} 
\begin{tabular}{|p{3.5cm}|p{4.2cm}|}  
\hline
\textbf{Platform Parameters} & \textbf{Values} \\
\hline
Virtual channels per port & 4 \\
\hline
Flow control & Credit-based \\
\hline
Frequency & 2 GHz \\
\hline
Flit size & 128 bits \\
\hline
L1D Cache size & 64 KB \\
\hline
L2 Cache size & 2 MB \\
\hline
Memory size & 512 MB \\
\hline
Cacheline size & 64 B \\
\hline
Cache coherence protocol & Dirctory-based MESI \\
\hline
Topology types & Crossbar, Mesh, Pt2Pt, Torus, FatTree, FlattenedButterFly\\
\hline
\end{tabular}
\label{tab:setup}
\end{table}


All CCTA traces extracted from Gem5’s output files are utilized as the state representations for the learning environment. The training process is structured into multiple episodes, where each episode corresponds to the execution of a complete application workload. At each decision step within an episode, the current state is passed to the Q-network, which evaluates the Q-values for candidate network topologies—including Crossbar, Mesh, Pt2Pt, Torus, Fat Tree, and Flattened Butterfly. An $\epsilon$-greedy policy then selects one of these topologies and the Link Weight Update module (a three-layer MLP) refines routing. The updated configuration is simulated in the Gem5 environment, which produces updated metrics for NoC traffic and cache coherence behavior. Based on these results, a multi-objective reward is computed that drives backpropagation for both the Q-network and the MLP-based refinement module, with the state updated before the next step. Although experiments focus on one application at a time, the framework can handle multiple workloads by randomly selecting benchmarks per episode while leaving the rest of the training process unchanged.

We compare methods that account for cache coherence with those that do not. For the latter, we evaluate dimension-order routing (XY), Q-learning-based methods like BiLCQ \cite{b31} and CrQ \cite{b32}, as well as the congestion-adaptive DyAD \cite{b33}. While BiLCQ and CrQ use Q-tables for updates, DyAD makes decisions based on queue lengths without Q-learning. To ensure a fair comparison, we integrate our Cache Coherence Traffic Analyzer (CCTA) with RL-based \cite{b29} and DRL-based \cite{b30} methods(i.e., non-cc-aware RL and non-cc-aware DRL) transforming them into cache coherence-aware versions, referring to as cc-aware RL and cc-aware DRL. This enables direct comparison within a cache-coherent environment. 

\subsection{Analysis of Cache Coherence Necessity}
\begin{figure}[h!]
    \centering
    \includegraphics[width=\linewidth]{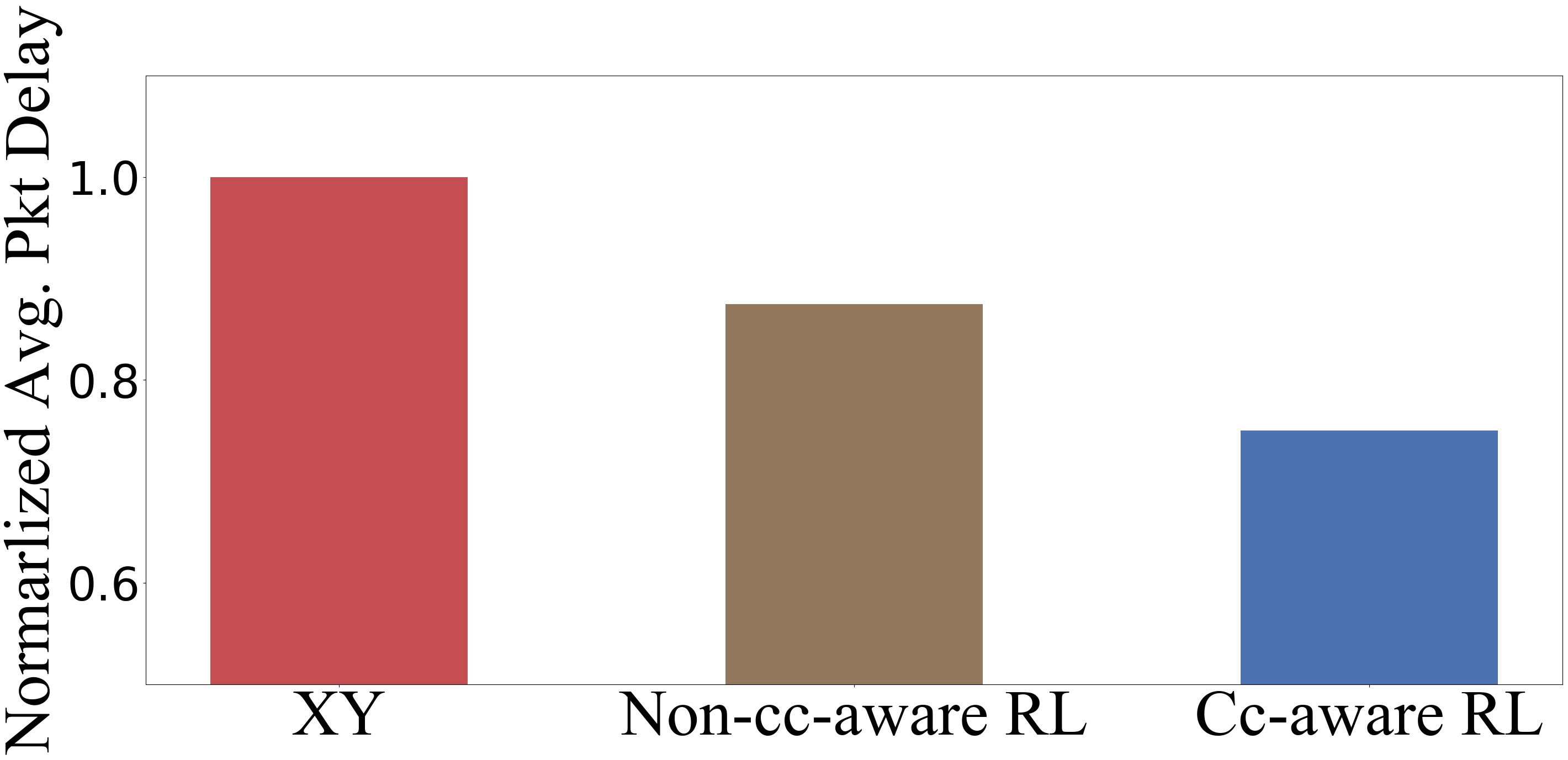}
    \caption{Normalized Average Packet Delay for PARSEC.}
    \label{fig:average packet delay} 
\end{figure}

To assess the necessity of incorporating cache coherence in routing, we compare XY, non-cc-aware RL, and cc-aware RL based on average packet delay. As shown in Figure \ref{fig:average packet delay}, cc-aware RL reduces packet delay by up to 17.65\% and 14.29\% compared to XY and non-cc-aware RL. Since both RL-based methods share the same routing approach, the reduction in delay highlights the role of cache coherence in improving data transmission efficiency. Furthermore, the results confirm that CCTA accurately captures cache coherence performance without compromising evaluation accuracy.

\subsection{Analysis of Cache Coherence Impact}

This section examines how cache coherence affects NoC traffic. First, we compare NoC energy consumption for non-cc-aware RL, non-cc-aware DRL, and our routing framework without cache coherence, showing our routing method’s superior efficiency. Next, we evaluate average packet latency, total energy consumption, and execution time in both cache-coherent and non-cache-coherent scenarios, highlighting the influence of cache coherence on NoC design and the effectiveness of our approach.

\begin{figure}[h!]
    \centering
    \includegraphics[width=\linewidth]{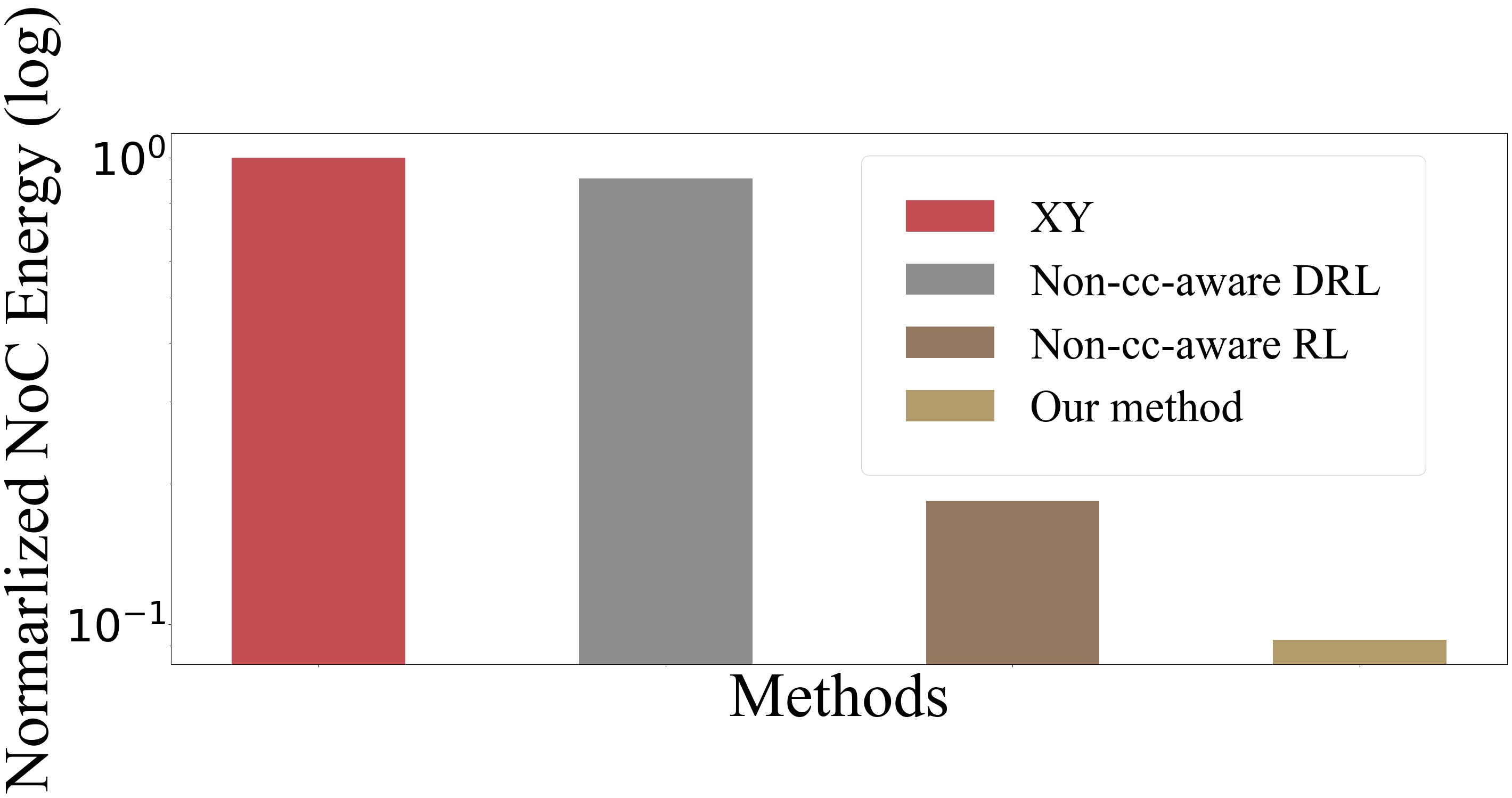}
    \caption{Normalized Total NoC Energy for PARSEC.}
    \label{fig:Total NoC energy} 
\end{figure}

\textbf{Results on NoC Energy.} As illustrated in Figure \ref{fig:Total NoC energy}, our DRL-based routing framework (without cache coherence consideration) reduces total NoC energy consumption by up to 90.75\%, 89.75\%, and 49.68\% compared to XY, non-cc-aware RL, and non-cc-aware DRL, respectively, demonstrating its efficiency in optimizing both path selection and traffic management. 

\begin{figure}[htbp]
    \centering
    
    \begin{subfigure}[b]{0.50\textwidth}
        \centering
        \includegraphics[width=\linewidth]{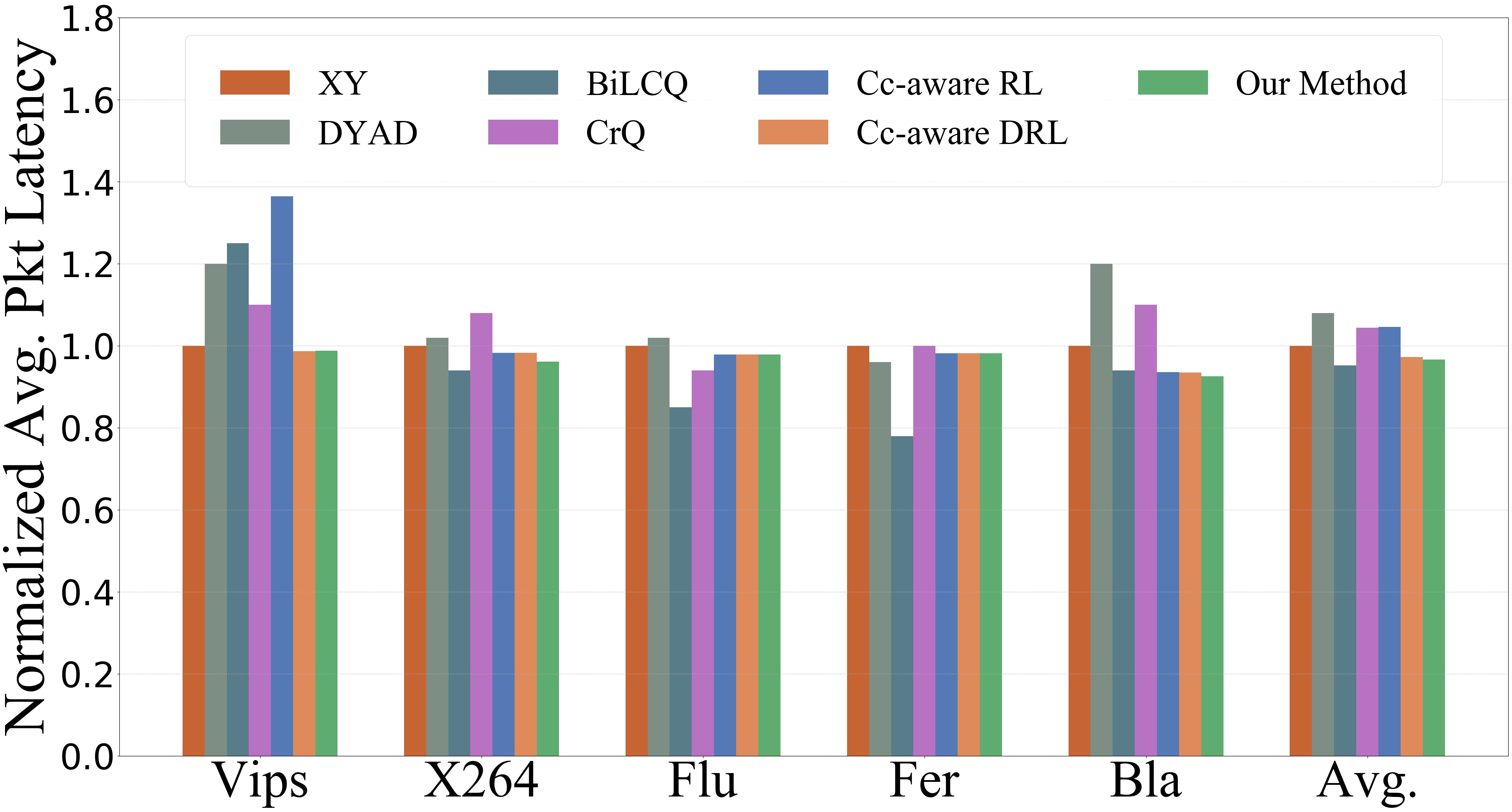}
        \caption{Average packet latency}
        \label{fig:Average packet latency}
    \end{subfigure}

    \begin{subfigure}[b]{0.50\textwidth}  
        \centering
        \includegraphics[width=\linewidth]{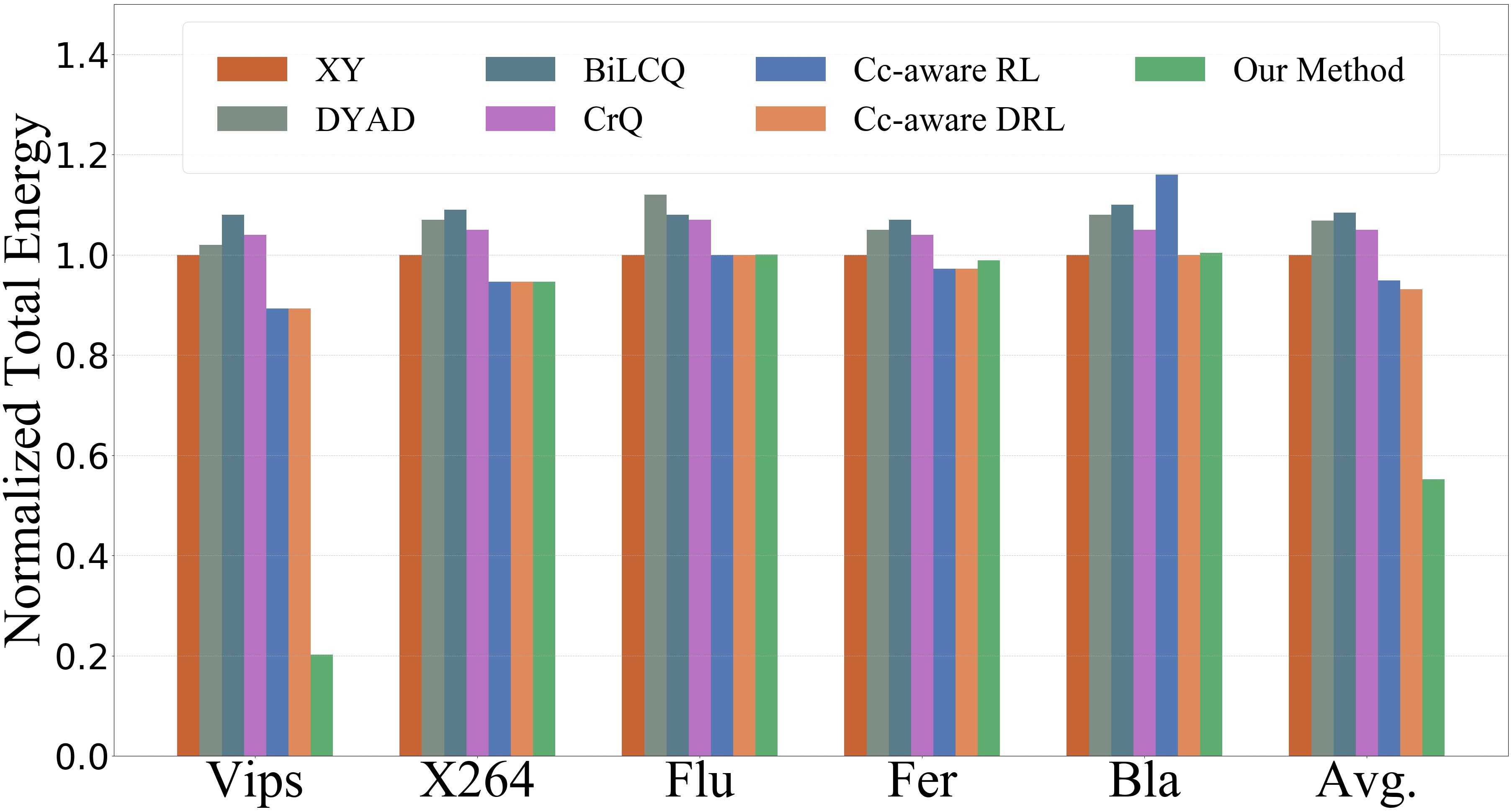}
        \caption{Total energy}
        \label{fig:Total energy}
    \end{subfigure}  

    \begin{subfigure}[b]{0.50\textwidth}
        \centering
        \includegraphics[width=\linewidth]{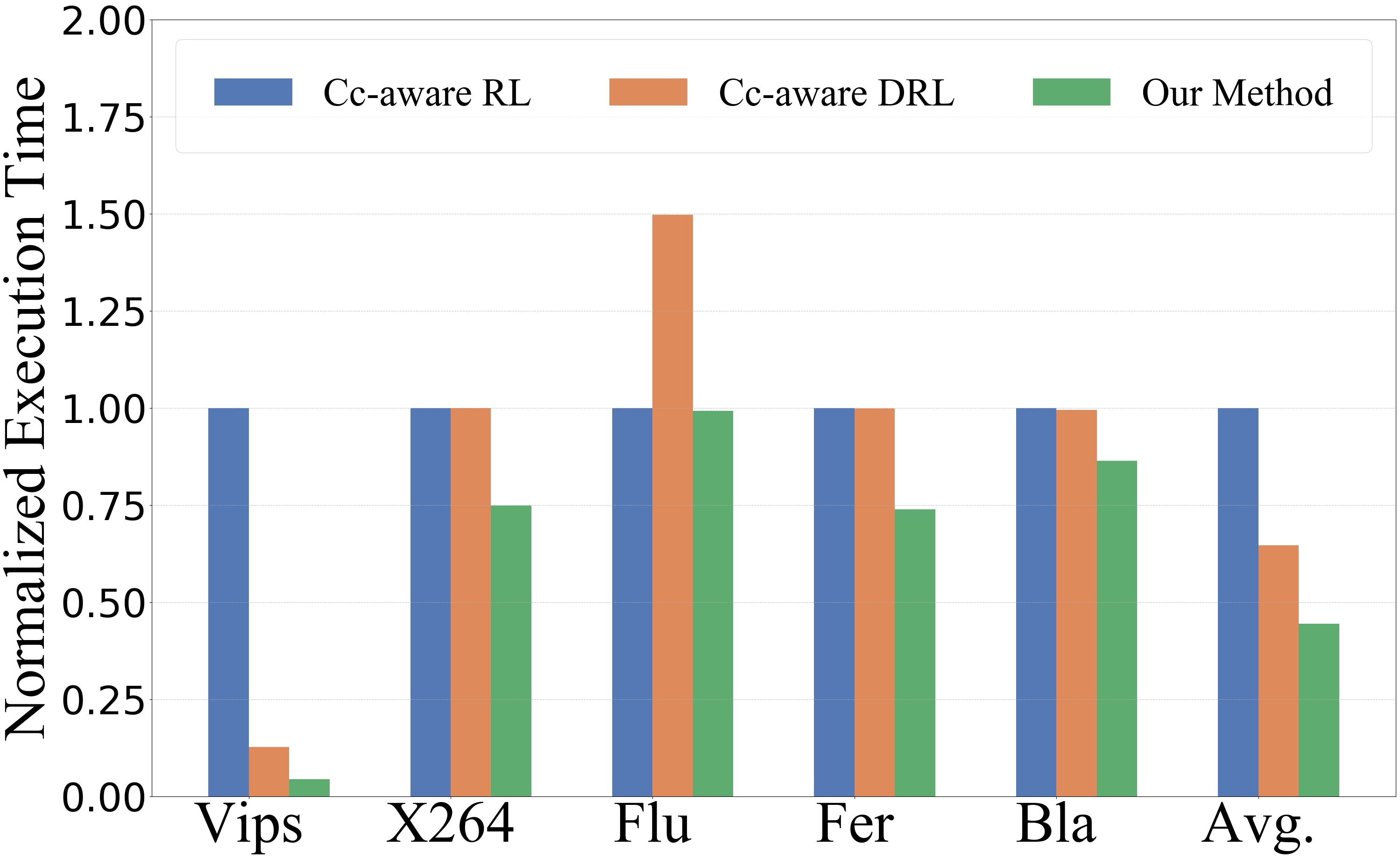}
        \caption{Execution time}
        \label{fig:execution time}
    \end{subfigure}
      
    \caption{Normalized performance in PARSEC: (a)Average packet latency, (b) Total energy, and (c) Execution time.}
    \label{fig:effeciency of CCTA and routing.}
\end{figure}

\begin{figure}[h!]
    \centering
    \includegraphics[width=\linewidth]{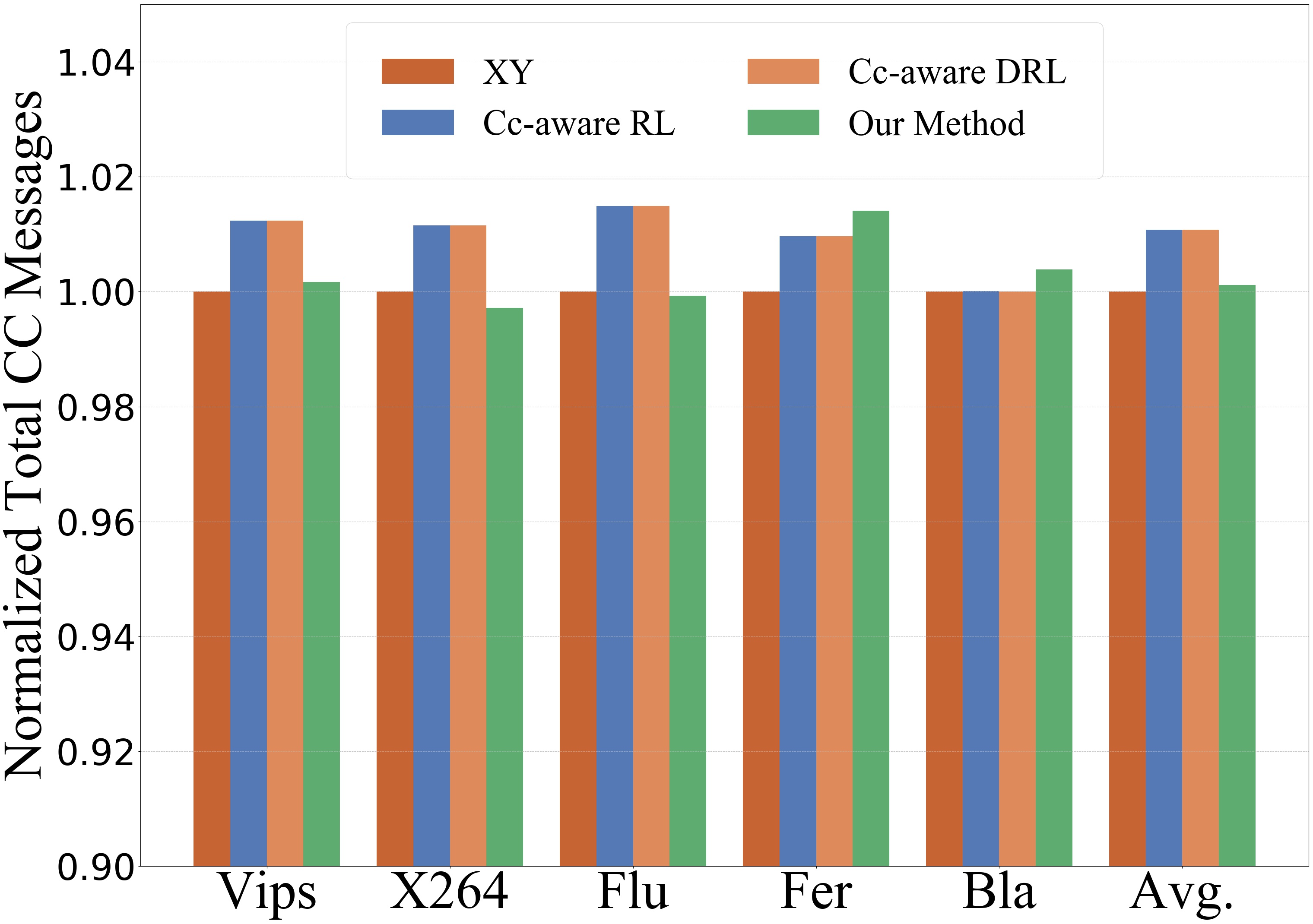}
    \caption{Normalized total messages driven by cache coherence protocol for PARSEC.}
    \label{fig:cc messages} 
\end{figure}
\begin{figure}[h!]
    \centering
    \includegraphics[width=\linewidth]{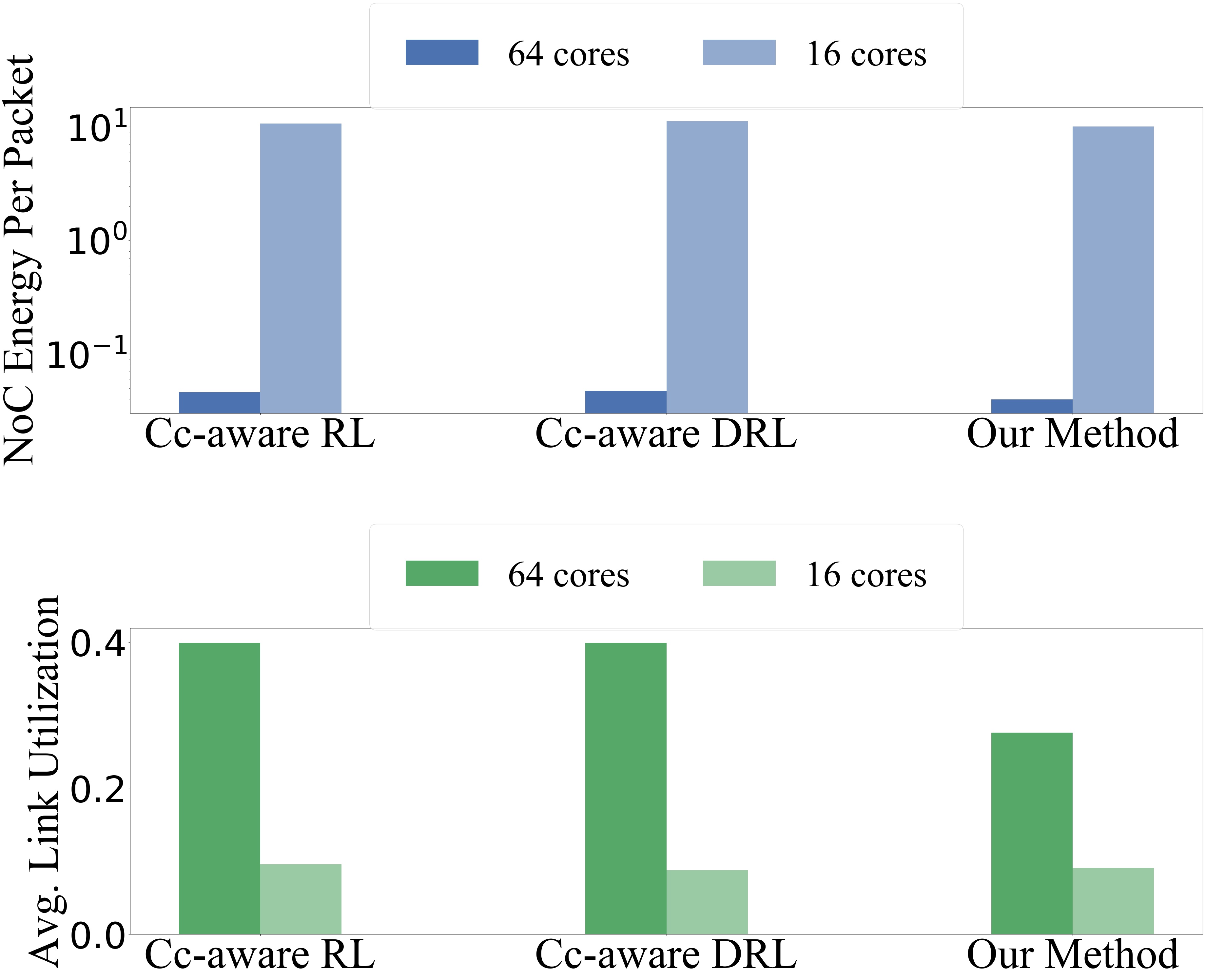}
    \caption{Comparison of total NoC energy per packet and average link utilization across 16 and 64 cores.}
    \label{fig:16_64_comparison} 
\end{figure}

\textbf{Results on Average Packet Latency.} Figure \ref{fig:Average packet latency} shows that our method consistently outperforms other routing strategies, reducing average latency by 3.36\%, 10.52\%, 7.43\%, 7.65\%, and 0.63\% compared to XY, DyAD, CrQ, cc-aware RL-based, and cc-aware DRL-based methods, respectively, while incurring 1.51\% more latency than BiLCQ. Note that XY, DyAD, CrQ, and BiLCQ operate without cache coherence. For workloads such as X264, Flu, and Fer, BiLCQ achieves lower latency, as it employs the RL-based routing which only optimizes latency without considering cache coherence or related overhead, it can aggressively reduce delays in some workloads but at the expense of increased overall energy usage (Figure \ref{fig:Total energy}). In contrast, our method achieves a better trade-off by maintaining competitive latency while substantially reducing total energy consumption, enhancing overall system performance and greater long-term sustainability.

\textbf{Results on Total Energy.} Figure \ref{fig:Total energy} shows our method reduces energy consumption by 44.75\%, 48.26\%, 49.02\%, 47.38\%, 41.77\%, and 40.67\% relative to XY, DyAD, BiLCQ, CrQ, cc-aware RL, and cc-aware DRL. Non-cc-aware techniques consistently consume more energy. Under cache coherence, cc-aware RL and DRL match our energy and latency on all workloads except Vips by effectively managing congestion and coherence traffic. However, they rely on inflexible topologies and incremental updates, limiting synergy between topology selection, routing, and coherence. Our method dynamically integrates these elements, avoiding unnecessary overhead and achieving notably shorter execution time (Figure \ref{fig:execution time}).

\textbf{Results on Execution Time.}
Our method (Figure \ref{fig:execution time}) reduces execution time by 55.51\% and 31.20\% compared to cc-aware RL and cc-aware DRL, respectively. Notably, cc-aware RL and DRL achieve nearly identical execution times on X264, Fer, and Bla—workloads whose predictable, loop-driven communication phases allow both methods to incur comparable overhead and performance. In contrast, Flu and Vips exhibit more bursty, dynamic data-sharing patterns, causing the two methods to diverge in traffic management and produce greater variation in execution times.

\textbf{Results on Cache Coherence Performance.} Figure \ref{fig:cc messages} shows our method reduces coherence-induced message transmissions by 8.33\% and 0.95\% compared to cc-aware RL and DRL, respectively, while remaining stable across all workloads. Although coherence protocols inherently increase traffic, our proactive approach throttles congestion and boosts communication efficiency.For workloads like Fer and Bla, this approach slightly increases coherence messages to ensure stable data exchange and balanced packet distribution. Coupled with improved performance (Figure \ref{fig:effeciency of CCTA and routing.}), our method delivers a superior balance of NoC efficiency, coherence traffic, and overall system performance. Moreover, RL- and DRL-based routing show similar performance under cache coherence for certain workloads, underscoring the need for more advanced cc-aware frameworks like ours to handle coherence more effectively.

\subsection{Analysis of Scalability}
We compare total NoC energy per packet and average link utilization in 16-core and 64-core systems. As shown in Figure \ref{fig:16_64_comparison}, our method reduces NoC energy per packet by 16.03\% and 13.67\% compared to cc-aware RL and cc-aware DRL in the 16-core system, while boosting average link utilization by 30.82\%. In the 64-core system, it achieves even greater gains—10.61\% and 14.0\% higher link utilization and 25.19\% and 20.37\% lower energy per packet than cc-aware RL and cc-aware DRL, respectively—demonstrating superior scalability and efficiency as core counts grow.

\section{Conclusion}
\label{sec: conclusion}
In this work, we demonstrate that cache coherence significantly reduces task computation time by enabling efficient data sharing among caches, thereby enhancing overall system performance. Building on this insight,  we identify and address three critical gaps in current NoC design: (i) misalignment between design objectives and actual performance when considering cache coherence, (ii) lack of analysis tools, limiting the understanding of cache coherence’s impact, and (iii) unexploited potential of integrating topology selection into routing decisions. To address these challenges, we propose a DRL-based framework that jointly optimizes topology selection and routing by dynamically learning from both NoC and cache coherence metrics. Additionally, we develop a Gem5-compatible Cache Coherence Traffic Analyzer (CCTA) to enable precise, fine-grained evaluation of coherence behaviors, providing deeper insights into its role in NoC design. Our approach achieves up to 10.52\% lower packet latency, 55.51\% faster execution time, and 49.02\% total energy savings, while maintaining minimal coherence overhead. These results highlight the importance of explicitly considering cache coherence in NoC design, paving the way for NoC–coherence co-design and demonstrating the broad potential of our methodology.

\begin{acks}
This work is partially supported by the Ministry of Education, Singapore, under its Academic Research Fund Tier 1 (RG94/23). 
\end{acks}


\end{document}